\documentclass[twocolumn,secnumarabic,amssymb, nobibnotes, aps, pre]{revtex4-2}

\usepackage{framed}
\usepackage{amsfonts}
\usepackage{amsmath}
\usepackage{graphicx}
\usepackage{color}

\usepackage[hidelinks]{hyperref}
\usepackage{xcolor}
\hypersetup{
	colorlinks,
	linkcolor={blue!80},
	citecolor={blue},
	urlcolor={blue!80!black}
}

\begin{document}
	%\maketitle
\title{Implications of tristability on localization phenomena: a necking bifurcation's tale}
	
\author{Edem Kossi Akakpo$^1$, Marc Haelterman$^1$, Francois Leo$^1$, and Pedro Parra-Rivas$^{2,1}$}
\email{Corresponding author: pedro.parra-rivas@uniroma1.it}

\affiliation{$^1$ OPERA-photonics, Universit\'e libre de Bruxelles, 50 Avenue F. D. Roosevelt, CP 194/5, B-1050 Bruxelles, Belgium\\
	$^2$Dipartimento  di  Ingegneria  dell’Informazione, Elettronica  e  Telecomunicazioni,
	Sapienza  Universit{\'a}  di  Roma, via  Eudossiana  18, 00184  Rome, Italy\\
}

%	\pacs{42.65.-k, 05.45.Jn, 05.45.Vx, 05.45.Xt, 85.60.-q}

\begin{abstract}
	We analyze the implication of tristability on localization phenomena in one-dimensional extended dissipative systems. In this context, localized states appear due to the interaction and locking of front waves connecting different extended states. In the tristable regime investigated here two extended uniform states coexist with one periodic Turing pattern. This scenario leads to the transition from the standard-homoclinic-snaking-related localized states associated with uniform-pattern bistability to the collapsed-homoclinic-snaking-related states which arise in a uniform-bistable configuration. We find that this transition is mediated by the emergence of hybrid states through codimension-two necking bifurcations. To perform this study we use bifurcation analysis on a non-variational mean-field model describing the spatiotemporal dynamics of light pulses in passive Kerr cavities.  
\end{abstract}

\maketitle

\section{Introduction}
The formation of localized states, hereafter LSs, in one-dimensional extended systems which are far from thermodynamic equilibrium is closely related to the concept of front wave interaction and locking \cite{pomeau_front_1986,coullet_nature_1987,thual_localized_1988,coullet_localized_2002}. Front waves, also known as domain walls or switching waves, consist in smooth interphases connecting different extended, and coexisting, states such that the nature and morphology of the former are determined by the latter. We can mainly classify fronts in two groups: those connecting two uniform, or homogeneous, states, that we will call {\it uniform fronts}, and those appearing as interphases between uniform and non-uniform, or patterned, states, which we refer to as {\it patterned front}.

Independently of the group, two fronts with different polarities (e.g., imagine a kink and anti-kink in the uniform scenario) may lock to one another leading to the formation of LSs with different extensions (i.e., widths) and shapes. However, the nature of the fronts involved in such interaction determines the locking mechanism, and therefore the bifurcation structure, which in this context is known as {\it homoclinic snaking} \cite{woods_heteroclinic_1999, knobloch_homoclinic_2005}.

In a uniform bistable configuration, i.e., when the coexisting states are uniform, the LSs that emerge correspond to an isolated plateau of one uniform state embedded in the other one, and undergo the so-called {\it collapsed homoclinic snaking} \cite{knobloch_homoclinic_2005,parra-rivas_dark_2016}. When the interaction involves patterned fronts, the LSs formation, much more complex in this case, is related to a different organization known as {\it standard homoclinic snaking} \cite{woods_heteroclinic_1999,gomila_bifurcation_2007}.

These bifurcation structures are quite well understood in a bistable configuration, i.e., when only two extended states coexist within the same parameter range, and have been widely studied in far apart scientific disciplines such as nonlinear optics \cite{gomila_bifurcation_2007,parra-rivas_bifurcation_2018,parra-rivas_localized_2019,parra-rivas_parametric_2020,parra-rivas_origin_2021}, reaction-diffusion systems \cite{al_saadi_unified_2021,zelnik_implications_2018,al_saadi_transitions_2023}, or neuroscience \cite{schmidt_bumps_2020}, to only cite a few. 

Things can be quite different when more than two extended states coexist. In this case, more complex hybrid fronts may appear, leading to important implications regarding the formation of new and exotic LSs. The implications of uniform-pattern-uniform tristability on the formation and bifurcation structure of LSs was tackled in a pant ecology context using Klausmeier-Gray-Scott model \cite{zelnik_implications_2018}. However, although a transition between the previous snaking organizations was found, the main mechanism, in bifurcation terms, behind such metamorphosis was veiled. This mechanism has been recently identified  in the prototypical, and variational, Swift-Hohenberg equation, and relies on the occurrence of a cascade of codimension-two (codim-2) {\it necking bifurcations} \cite{parra-rivas_organization_2023}. In these codim-2 points, two fold bifurcations collide and separate afterward, yielding the reorganization of the LSs bifurcation structures. Despite this discovery, the generality of these results, and therefore their extension, to other scenarios, particularly  non-variational ones, was not demonstrated.  

In this regard, the main scope of this work is to prove such generality. To do so, we use a modified version of the Lugiato-Lefever equation (LLE), a non-variational model (i.e., with nongradient-like dynamics \cite{cross_pattern_1993}) describing the evolution of the complex amplitude of the electric field inside a passive Kerr optical cavitiy \cite{lugiato_spatial_1987, haelterman_dissipative_1992}. Besides the importance of understanding this transition from a fundamental point of view, this knowledge will have important implications from a more practical perspective since it could provide new ways for optical cavity soliton and frequency comb generation \cite{rowley_self-emergence_2022,lottes_excitation_2021}. 

The paper is organized as follows. In Section~\ref{sec:2} we introduce a modified version of the LLE equation describing Kerr passive cavities in the presence of fourth-order dispersion (FOD). This dispersion effect is essential for the emergence of tristability. Later, in Section~\ref{sec:3} we introduce the necessary background for understanding stability properties of the uniform and extended states and present the tristable scenario. Section~\ref{sec:4} unveils the LSs bifurcation structure transition taking place in the system. The necking bifurcation organizing these transitions is presented in Section~\ref{sec:5}. Finally, in Section~\ref{sec:6} we provide a short discussion and the main conclusions of our work.  
\section{Model and methods}\label{sec:2}
In the mean-field approximation, passive Kerr dispersive cavities can be described by the LLE \cite{haelterman_dissipative_1992, chembo_spatiotemporal_2013}. Considering chromatic dispersion up to fourth-order, and neglecting the contribution of third-order dispersion (TOD), the normalized LLE reads
\begin{equation}\label{LLE}
	\partial_t A = -(1 + i \Delta) A - i d_2 \partial_x^2 A + i d_4 \partial_x^4 A + i |A|^2 A + S,
\end{equation}
where $A$ is the complex field amplitude, $t$ represents the time coordinate, and $x$ the space in spatial cavities, the fast time in fiber cavities or angular variable in microresonators. The losses are normalized to 1, $\Delta$ is the phase detuning from the closest cavity resonance and $S$ in the driving field amplitude. With this normalization,  $d_2={\rm sign}(\beta_2)=\pm 1$ and $d_4\equiv\beta_4 \alpha/(6L|\beta_2|^2)$, where $\alpha$ represents the losses, $L$ is the cavity length, and $\beta_2$ and $\beta_4$ represent the second-order dispersion (SOD) and FOD coefficients, respectively.

In this work we are interested in studying LSs which are symmetric through the transformation $x\rightarrow -x$. This is the reason why we have neglected TOD effects. Besides, as we will see in the next section, the presence of FOD is necessary for the emergence of uniform-pattern-uniform tristability, which is otherwise absent \cite{parra-rivas_origin_2021}. 
To attain such tristable configuration we fix $d_2=-1$ and $d_4=1$. This regime
was investigated in 2010 when the positive effect of FOD for stabilizing standard-homoclinic-snaking-related dark LSs was identified \cite{tlidi_high-order_2010}.
We have recently used Eq.~(\ref{LLE}) for studying the formation of collapsed-snaking-related LSs, but using a rather different dispersion configuration (i.e., $d_2=1$   and $d_4>0$) \cite{akakpo_emergence_2023}. Moreover, this model has been also exploited for unveil the formation of bright and dark LSs in pure quartic dispersive cavities \cite{parra-rivas_quartic_2022}.

To perform this bifurcation analysis, we mainly apply path-continuation algorithms  
based on a predictor-corrector method \cite{doedel_numerical_1991,doedel_numerical_1991-1} using the free software package AUTO-07p \cite{Doedel2009}.   
With this procedure, we are able to compute not only the stable but also the unstable state solutions, unveiling their connection. The spectral stability of these states is then determined by solving the linear eigenvalue problem
\begin{equation}\label{lin}
	\mathcal{L}\psi=\sigma\psi,
\end{equation}
where $\mathcal{L}$ is the linear operator associated with the right hand side of Eq.~(\ref{LLE}) evaluated a given steady state, and $\psi$ is the eigenmode corresponding to the eigenvalue $\sigma$. The time-independent state is stable if Re$[\sigma]<0$, and unstable otherwise. If by varying a control parameter of the system, let's say $p$, this transition occurs at the value $p=p_c$ (i.e., Re$[\sigma(p_c)]=0$), we say that a local bifurcation takes place at $p_c$ \cite{wiggins_introduction_2003,scroggie_pattern_1994}.

For representing the different type of states and their bifurcation structure, we compute bifurcation and phase diagrams where we show the modification of a given bifurcation measure as a function of the main parameter of the system. Here, we choose as measure the solutions energy (i.e., the $L_2$-norm) $$E^2\equiv L^{-1}\int_{-L/2}^{L/2}|A(x)|^2 dx,$$
and $S$ and $\Delta$ as main control parameters of the system.

\section{Tristability and extended steady states}\label{sec:3}
\begin{figure*}[!t]
	\centering
	\includegraphics[scale=1]{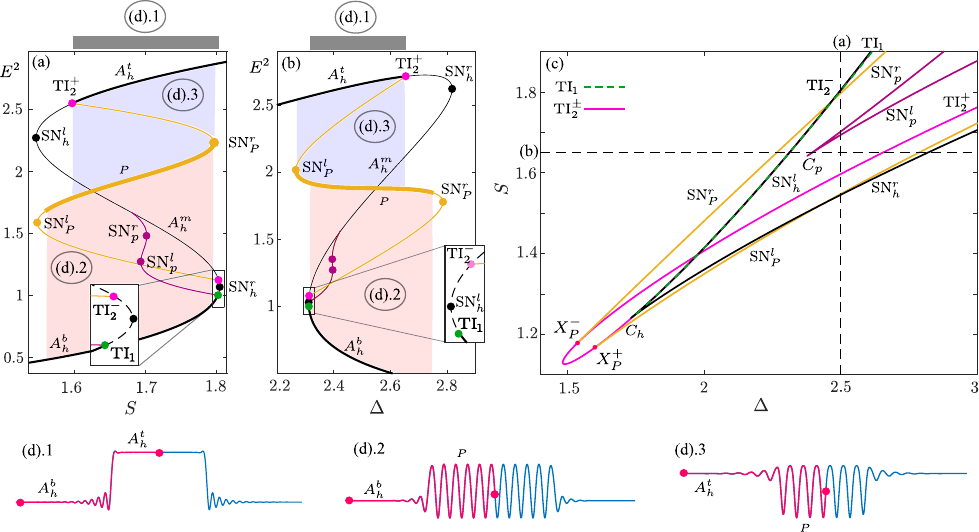}
	\caption{\textbf{Uniform-pattern-uniform tristable regime.} Panel (a) shows the modification of the uniform state $A_h$ (see black curve) and the Turing pattern $P$ (see yellow curve) energy $E^2$ as a function of $S$ for $\Delta=2.5$. The gray bar on the top defines the region of uniform bistability. The blue and red shaded colored areas correspond to uniform-pattern bistability. Panel (b) shows the nonlinear cavity resonance for $S=1.65$. In both cases we mark the saddle-node bifurcations of the pattern SN$_P^{l,r}$, those of the uniform state SN$_h^{l,r}$, and the Turing bifurcations TI$_1$ and TI$_2^\pm$. Stable (unstable) are illustrated using solid thick (thin) lines.
		Panel (c) shows the modifications of SN$_h^{l,r}$, SN$_P^{l,r}$, TI$_1$ and TI$_2^\pm$ in the $(\Delta,S)$-parameter space.  The vertical dashed line corresponds to the situation shown in (a); the horizontal one to the nonlinear cavity resonance shown in (b). We also mark the codim-2 points $C_h$ and $X_P^\pm$. In (d) we show some examples of the type of LSs (blue) and front waves (pink) that can form in the bistable scenarios shown in (a) and (b).  }
	\label{fig1}
\end{figure*}
When working with dissipative extended systems, one essential task is to identify the different types of extended time-independent states and their stability. 
In this regard, the simplest time-independent solution is the uniform or homogeneous steady state (HSS) solution $A_h$, which satisfies the algebraic equation  \cite{lugiato_spatial_1987,haelterman_dissipative_1992} 
%\begin{equation}
%	(1 + i \Delta) A_h - i |A_h|^2 A_h - S=0.
%\end{equation}
%This equation can be rewritten in the form 
\begin{equation}\label{Sshape}
	S^2 = I_h^3 - 2\Delta I_h^2 + (1 + \Delta^2)I_h,
\end{equation}
for $I_h\equiv|A_h|^2$. For a fixed value of $\Delta$, this expression defines a nonlinear dependence between the intracavity intensity $I_h$ and the pump $S$ which can be seen in Fig.~\ref{fig1}(a) for $\Delta=2.5$. In this regime, the system shows uniform multistability, i.e., for the same value of $S$ three HSSs, namely $A_h^b$, $A_h^m$, and $A_h^t$, coexist. These states are connected at two folds, or turning points, located at the positions 
\begin{equation}\label{fold_points}
	I_h^{l,r}\equiv\frac{1}{3}\left(2\Delta\pm\sqrt{\Delta^2-3}\right).
\end{equation}
As a function of $\Delta$, these folds define two curves in the $(\Delta,S)$-parameter diagram shown in Fig.~\ref{fig1}(c) [see solid black lines]. Here, the vertical pointed line corresponds to the diagram plotted in Fig.~\ref{fig1}(a). Decreasing $\Delta$, eventually, these folds annhilate one another in a codim-2 cusp bifurcation, marked with $C_h$, which occurs at $\Delta=\sqrt{3}$. Similarly, one can look at this scenario by fixing $S$ and varying $\Delta$. As result, one obtains the nonlinear resonance of the cavity which is solution of the equation $\Delta=I_h\pm\sqrt{(S/I_h)^2-1}$. This resonance is plotted in Fig.~\ref{fig1}(b) for $S=1.65$. 

To establish the stability of these uniform states we perform a linear stability analysis by studying how $A_h$ reacts against non-uniform perturbations of the form $\propto e^{\sigma t+ikx}$. This leads to the linear eigenvalue problem (\ref{lin}). For $k=0$, $A_h$ undergoes a pair of saddle-node bifurcations SN$_h^{l,r}$ whose positions are given by the folds Eq.~(\ref{fold_points}). For non-uniform perturbations ($k\neq0$), $A_h$ encounters a pair of Turing instabilities which are marked TI$_1$ and TI$_2$ in in Fig.~\ref{fig1}. These instabilities appear at different positions and have different critical wave-numbers. TI$_1$ [see green dot in the close-up view of Figs.~\ref{fig1}(a),(b)] occurs, for $\Delta>7/4$, at $I_h=I_1^c\equiv1$ with wave-number 
%\begin{equation}
%			k_{1}^c= \pm \sqrt{\frac{-d_2 \pm \sqrt{1 - 4 d_4 (2  - \Delta)}}{2 d_4}}.
%\end{equation}
\begin{equation}
	k_{1}^c= \sqrt{\frac{1+\sqrt{1 + 4(\Delta-2)}}{2}}.
\end{equation}
The wave-number and position associated with TI$_2$ (pink) read
% $k^c_{2} =\sqrt{-d_2/(2 d_4)}$ 
 $k^c_{2} =1/\sqrt{2}$  
 and 
\begin{equation*}
 		%I_{2}^c = \frac{ 4 \Delta d_4 + 1}{6d_4}\pm \frac{\sqrt{16 \Delta^2 d_4^2 + 8 \Delta d_4-48 d_4^2+1}}{12 d_4},	
I_{2}^c = \frac{ 4 \Delta + 1}{6}\pm \frac{\sqrt{(\Delta+1/4)^2-3}}{3},
%I_{2}^c = \frac{ 4 \Delta + 1}{6}\pm \frac{\sqrt{16 \Delta^2+8\Delta-48+1}}{12},
\end{equation*}
with $\Delta>\sqrt{3}-1/4$. The parts of the $A_h$-solution curves which are Turing unstable are depicted using thin solid lines, while for those which are stable we use solid thick lines. The modification of these bifurcations with varying $\Delta$ are shown in Fig.~\ref{fig1}(c) using a pink line for TI$_2$ and a green dashed line for TI$_1$. At this point we can clearly establish the uniform-bistability region where $A_h^{b,t}$ coexist and are both stable. This regions is depicted using a gray shaded box on top of Figs.~\ref{fig1}(a),(b).
\begin{figure*}[!t]
	\centering
	\includegraphics[scale=1]{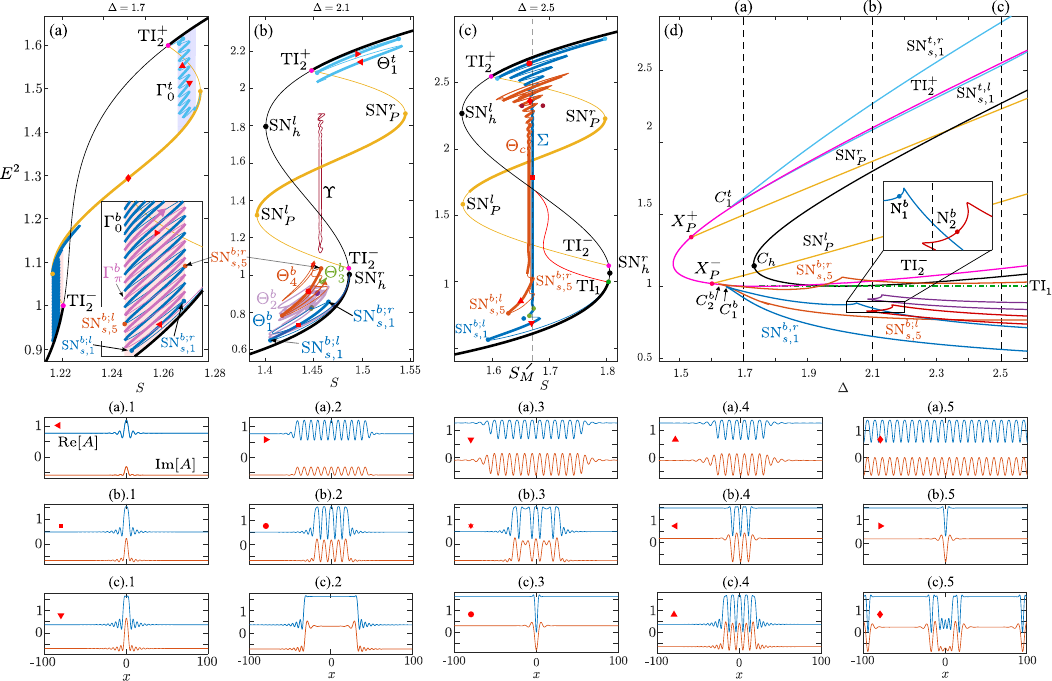}
	\caption{\textbf{Transition between the standard and collapsed homoclinic snaking in a uniform-pattern-uniform tristable regime.} Panel (a) shows the standard homoclinic regime for $\Delta=1.7$. The bifurcation diagrams  $\Gamma^b_{0,\pi}$ and $\Gamma^t$ correspond to bright and dark LSs, respectively. The yellow curve represents the Turing pattern and the red- and blue-shaded boxes represent the uniform-pattern bistability regions. Panels (a).1-5 illustrate examples of LSs along these diagrams. Panel (b) corresponds to the transition region and shows the emerging isolas $\Theta^{b}_{1,2,3,4}$, $\Theta^t_1$ and $\Upsilon$. LSs on these curves are plotted in panels (b).1-5. Panel (c) represents the collapse snaking region (see $\Sigma$ diagram), where other complex isolas also exist (see $\Theta_c$). Examples of states belonging to these branches are depicted Figs.~\ref{fig2}(c).1-5. Panel (d) shows the modification of the main bifurcation lines as a function of $E^2$ and $\Delta$.  }
	\label{fig2}
\end{figure*}
From TI$_2$, a spatially periodic Turing pattern $P$ emerges with a wave-number which is approximately equal to $k^c_{2}$. For the parameter values we choose here, this pattern arises subcritically from $I_2^{c\pm}$ [see yellow curve Fig.~\ref{fig1}(a)] and stabilizes at the SN bifurcations SN$_P^{l,r}$. The modification of these folds in the $(\Delta,S)$-parameter space is depicted in Fig.~\ref{fig1}(c). With decreasing $\Delta$, this Turing pattern becomes supercritical at the codim-2 points $X_P^{\pm}$, which correspond to degenerate Turing instabilities, from where the supercritical and subcritical configurations arise. % i.e., to the transition between super and subcritical . 

Regarding TI$_1$, another unstable subcritical Turing pattern arise from $I_1^c$, now with a typical wavelength $2\pi/k_1^c$ [see magenta curve in Figs.~\ref{fig1}(a),(b)]. This state undergoes a pair of fold bifurcations SN$_p^{l,r}$ that leave their stability unchanged, i.e., this pattern never become stable. With increasing $\Delta$, SN$_p^{l,r}$ separate from one another. Decreasing  $\Delta$, however, they come closer and eventually they annihilate each other in the cusp bifurcation $C_p$ [see magenta lines in Figs.~\ref{fig1}(c)]. This pattern always arise subcritically from $A_h^b$. 
 
The stable branch of the $P$-pattern [see tick solid line in Figs.~\ref{fig1}(a),(b)] and stable sections of $A_h$ determine a pair of pattern-uniform bistability regions colored using shaded blue and red [see Figs.~\ref{fig1}(a),(b)]. These regions, together with the uniform-bistability interval marked with the gray bar, form a tristable uniform-pattern-uniform scenario. In this context [see Fig.~\ref{fig1}(d)], different front states (see pink profile) and LSs (see blue profile) may form. The existence of these states in the different parameter regions of the systems are marked in Figs.~\ref{fig1}(a),(b). 
In 1D extended systems, the interaction and locking of these fronts is well understood and we refer to the following works for more details \cite{woods_heteroclinic_1999,coullet_localized_2002,gomila_bifurcation_2007, makrides_existence_2019,parra-rivas_origin_2021}.

The coexistence between all these bistable configurations changes with $\Delta$: decreasing $\Delta$, uniform bistability disappears at $C_h$ and only the upper and bottom uniform-pattern bistable regimes persist until reaching $X_{P}^\pm$; above $C_h$ uniform-pattern-uniform tristability emerges, increasing its extension with $\Delta$.

\section{Bifurcation structure transition in tristable regimes}\label{sec:4}

Once the uniform and pattern states have been investigated, and their overlapping regions and variety of fronts identified, it is now time for studying the emergence of LSs. 

To do so, we perform a detailed and systematic bifurcation analysis based on path-parameter continuation techniques and spectral analysis, as already introduced in Section~\ref{sec:2}. The main results are depicted in Fig.~\ref{fig2} where the modification of the time-independent extended and localized states bifurcation structure is shown in terms of their energy $E^2$ for different values $\Delta$.

Figure~\ref{fig2}(a) depicts, for $\Delta=1.7$, the LSs bifurcation structure which arises through the locking of patterned fronts like those shown in Figs.~\ref{fig1}(d).2 and \ref{fig1}(d).3 in a pure uniform-pattern bistable regime. Here, the periodic pattern $P$ emerge subcritically from the Turing bifurcations TI$_2^\pm$ (see position $I_2^{c\pm}$ located at the bottom and top of the HSS diagram), creating two bistable regimes which are colored in shaded blue and red. Note that, for this value of $\Delta$, there is not uniform bistability and the uniform-pattern bistable regimes do not overlap. 

Within these bistable regimes, the locking of patterned-fronts yields the creation of dark and bright LSs which organize in terms of {\it standard homoclinic snaking} \cite{woods_heteroclinic_1999}. In this bifurcation structure the solution curves oscillates back-and-forth within the so called pinning region \cite{woods_heteroclinic_1999,gomila_bifurcation_2007,burke_snakes_2007}. Along this diagram, LSs consist in a slug of the periodic pattern [see Fig.~\ref{fig2}(a).5] embedded within a uniform background. Some examples of these states are illustrated in Figs.~\ref{fig2}(a).1-4.

These LSs emerge from TI$_2^\pm$ subcritically in pairs yielding two families of snaking curves that we call $\Gamma_{0,\pi}$ \cite{burke_snakes_2007,parra-rivas_origin_2021}. While $\Gamma_{0}$ corresponds to LSs with a odd number of pattern wave-lengths (i.e., peaks), in $\Gamma_{\pi}$ the number of peaks are even. These two families are shown for the bright LSs case [see $\Gamma^b_{0,\pi}$ in Fig.~\ref{fig2}(a)]. Regarding dark LSs we only plot $\Gamma^t_0$ for simplicity. Along these diagrams there is an alternation between stable and unstable states, which are separated by SN bifurcations that we label SN$^{l,r}_{s,i}$, with $i$ representing the numbers of pattern peaks within the LS solution branch. These left/right bifurcations limit two well defined LS-mustistability regions whose width is conserved as varying $E^2$ [see close-up view in Fig.~\ref{fig2}(a) as reference]. 
While the snaking associated with the bright states exists in the absence of $d_4$ \cite{gomila_bifurcation_2007,parra-rivas_origin_2021}, the dark ones are stabilized precisely due to the presence of this dispersive effect \cite{tlidi_high-order_2010}.

The modification of the main bifurcations of the system with $\Delta$, and therefore the modification of the LSs bifurcation structure, is reflected in the $(E^2,\Delta)$-phase diagram shown in Fig.~\ref{fig2}(d). The bifurcation diagram illustrated in Fig.~\ref{fig2}(a) corresponds to a slice of the phase diagram for $\Delta=1.7$ [see first vertical dashed line]. 

This diagram shows in yellow SN$_P^{l,r}$, in black SN$_h^{l,r}$ and in pink TI$_2^\pm$. The codim-2 point where the Turing patterns become subcritical are marked with the $X_{P}^\pm$ as in Fig.~\ref{fig1}(c). The bifurcations SN$_P^{l,r}$ emanate from this point. The cusp codim-2 point shown in Fig.~\ref{fig1}(c) here corresponds to the fold marked with $C_h$. Besides, SN$_{s,i}^{b;l,r}$ for $i=1,5$ and SN$_{s,1}^{t;l,r}$ are plotted. Here, the super-index $b$ (or $t$) refers to the bottom (top) part of the diagram shown in Fig.~\ref{fig2}(a).   Decreasing $\Delta$ these bifurcation pairs annhiliate one-another in different cusp points $C^{b,t}_i$, and with them their associated LSs. 

Just a bit after the emergence of the uniform-bistability (i.e., for $\Delta>\sqrt{3}$) the standard homoclinic snaking scenario breaks down, leading to the configuration depicted in Fig.~\ref{fig2}(b) [$\Delta=2.1$] where a sequence of isolas (i.e., disconnected closed bifurcation curves) is formed. The blue isola in the bottom, $\Theta^b_1$, corresponds to the single-peak LS depicted in Fig.~\ref{fig2}(b).1 and still arises from TI$_2^-$. The stable branch in this curve is the deformation of the first stable branch of $\Gamma_0^b$ [see Fig.~\ref{fig2}(a)]. Similarly occurs for all LSs, leading to an intricate entangle of isolas. Here, for simplicity, we also show the isolas $\Theta^b_{2,3}$ and $\Theta^b_{4}$. The latter is related to the five-peak state plotted in Fig.~\ref{fig2}(b).2 and to the {\it hybrid} LSs formed through a combination of the two locking mechanism described previously [see Fig.~\ref{fig2}(b).3].% One example of such a state is shown in Fig.~\ref{fig2}(b).3.

In a similar fashion, the homoclinic snaking $\Gamma_0^t$ gets disconnected forming another sequence of isolas. One of these isolas, $\Theta^t_1$, is depicted in Fig.~\ref{fig2}(b). The stable branches of $\Theta^t_1$ correspond to the dark LSs shown in Fig.~\ref{fig2}(b).4 and \ref{fig2}(b).5. 

In-between isolas $\Theta^b_{1,2,3,4}$ and $\Theta^t_1$, approximately in the middle of the uniform-pattern bistability region, a new type of isola ($\Upsilon$) appears.  All along this isola, LSs form through the locking of uniform-fronts. $\Upsilon$ is the main embryonic structure underlying the collapsed snaking formation.
% corresponds to LSs whose formation relies on a hybrid combination of the two locking mechanism described previously. . In what follows we will refer to these LSs as {\it hybrid} states. {\bf\color{blue}Not shown!}  

By increasing $\Delta$, $\Theta^b_{1,2,3}$ and $\Upsilon$ will merge leading to the recombination of the LSs branches forming $\Upsilon'$ [see Fig.~\ref{fig3}(b)], which eventually, will yield the appearance of the collapse homoclinic snaking $\Sigma$ shown in Fig.~\ref{fig2}(c) for $\Delta=2.5$ [see vertical dashed line in Fig.~\ref{fig2}(d)]. The emergence of this bifurcation structure is associated with the locking of uniform fronts [see Fig.~\ref{fig1}(d).1] and leads to a broad variety of dark and bright LSs [see Fig.~\ref{fig2}(c).1-3]. This bifurcation structure oscillates back and forth around the uniform Maxwell point ($S_M$) of the system in a damped fashion which is intrinsically related to the LSs formation in this regime \cite{parra-rivas_origin_2016,parra-rivas_origin_2021}. Here, in contrast to the standard case, LSs multistability decreases when moving away from $S_M$. The collapsed snaking exist in other operational regimes involving $d_4$ \cite{akakpo_emergence_2023}, and also in the absence of fourth order dispersion \cite{parra-rivas_origin_2021}. However, as far as we known, collapsed snaking has not been previously predicted in this regime.

Besides, isolas like $\Theta_c$, formed through the merging of $\Theta_4^b$ and other $\Upsilon$-like isolas [see Fig.~\ref{fig2}(b)], appear, combining states like the one shown in Fig.~\ref{fig2}(c).4 and Fig.~\ref{fig2}(c).5. Indeed, the hybrid state shown in Fig~\ref{fig2}(b).3 can be tracked by path-continuing the one depicted in Fig.~\ref{fig2}(c).4.

\section{Dissecting the transition: The necking bifurcation}\label{sec:5}
%------------------------------
%The destruction of the homoclinic snaking curves take place through the occurrence of a sequence of necking bifurcations as described in Ref.~\cite{parra-rivas_organization_2023}. 
%-----------------------------

\begin{figure}[!t]
	\centering
	\includegraphics[scale=1]{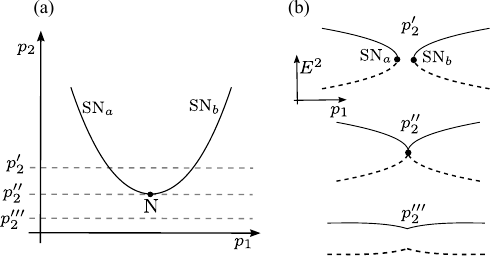}
	\caption{\textbf{The necking bifurcation unfolding.} (a) Localization of the saddle-node bifurations SN$_{a,b}$ in the parameter space $(p_1,p_2)$. In our case, we can consider $p_1=S$ and $p_2=\Delta$. The dashed lines indicate the fixed-$p_2$ slices depicted in (b) for $p=p_2',p_2''$ and $p_2'''$. The necking bifurcation N occurs at $p=p_2''$ when SN$_a$ and SN$_b$ collide.}
		 %shows the approaching ($p_2'$), collision ($p_2''$)  and  modification of SN$_{c,1}^b$ and SN$_{c,2}^b$, and the occurrence of the codim-2 bifurcations $C_{c,2}^l$, $C_{c,2}^r$ and the necking bifurcation N$_{c,2}^b$ in the $(\Delta,S)$-parameter space. The vertical dashed lines correspond to the diagrams shown in panel (b) and (c) for $\Delta=2.16$ and $\Delta=2.3$, respectively. Panel (b) shows the situation just below the occurrence of N$_{c,2}^b$: the isolas $\Theta_1^b$ and $\Upsilon'$ are very close to one another. Panel (c) illustrates the situation above N$_{2}^b$ after the collision of $\Theta_1^b$ and $\Upsilon'$ in N$_{2}^b$.}
	\label{fig3b}
\end{figure}

The questions that arise now is: how is this transition taking place? 
The destruction of one bifurcation structure and the reorganization of branches to form new ones is normally related with the merging and later splitting of solution branches occurring when modifying one or several control parameters of the system. Here, this re-connection takes place through a cascade of codim-2 {\it necking bifurcations} \cite{prat_12_2002}, hereafter N, like those shown in the close-up views of Fig.~\ref{fig2}(d) [see N$_{1,2}^b$].
In each N, a pair of saddle-node bifurcations belonging to different structures, e.g., one to a homoclinic snaking and another one to an isola, collide leading to the re-connection of solution branches and the reorganization of the bifurcation structure \cite{parra-rivas_organization_2023}. An schematic representation of the unfolding of this bifurcation is illustrated in Fig.~\ref{fig3b}.

\begin{figure}[!b]
	\centering
	\includegraphics[scale=1]{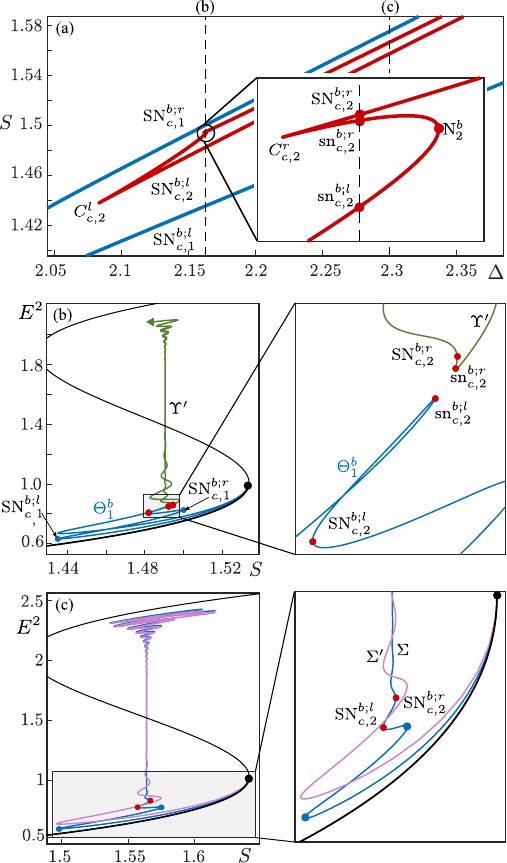}
	\caption{\textbf{The necking bifurcation in the collapsed snaking formation.} Panel (a) shows the modification of SN$_{c,1}^b$ and SN$_{c,2}^b$, and the occurrence of the codim-2 bifurcations $C_{c,2}^l$, $C_{c,2}^r$ and the necking bifurcation N$_{c,2}^b$ in the $(\Delta,S)$-parameter space. The vertical dashed lines correspond to the diagrams shown in panel (b) and (c) for $\Delta=2.16$ and $\Delta=2.3$, respectively. Panel (b) shows the situation just below the occurrence of N$_{c,2}^b$: the isolas $\Theta_1^b$ and $\Upsilon'$ are very close to one another. Panel (c) illustrates the situation above the collision of $\Theta_1^b$ and $\Upsilon'$ in N$_{2}^b$.}
	\label{fig3}
\end{figure}

To better understand the implication of this bifurcation in the snaking transition, let us take a closer look to its appearance near the collapsed snaking regime. Figure~\ref{fig3}(a) shows the modification of the first pair of saddle-node bifurcations SN$^{b;l,r}_{c,i}$ ($i=1,2$) in the $(\Delta,S)$-parameter space. The bifurcations SN$^{b;l,r}_{c,2}$ (see red curve) undergo the necking bifurcation N$_{2}^b$ which is shown in the close-up view. This bifurcation always appear 
together with a pair of cusp bifurcations that here we mark as $C_{c,2}^{l}$ and $C_{c,2}^{r}$.

Let us take a slice of Fig.~\ref{fig3}(a) for $\Delta=2.16$. The resulting bifurcation structure is shown in Fig.~\ref{fig3}(b). Here, two main isolas appear: $\Theta^b_1$ and $\Upsilon'$, both modifications of those already introduced in Fig.~\ref{fig2}(b). A close-up view of this diagram (see right panel) shows how these isolas are disconnected, although very close to one another. There we can see two pairs of saddle-node bifurcations: SN$_{c,2}^{b;l}$ and sn$_{c,2}^{b;l}$ which are created at $C_{c,2}^{l}$, and SN$_{c,2}^{b;r}$ and sn$_{c,2}^{b;r}$ that emerge from $C_{c,2}^{r}$ [see Fig.~\ref{fig3}(a)].  By increasing $\Delta$ further, sn$_{c,2}^{b;l}$ and sn$_{c,2}^{b;r}$ merge at N$_{2}^b$, leading to the re-connection of $\Theta^b_1$ and $\Upsilon'$. This process yields the situation depicted in Fig.~\ref{fig3}(c) for $\Delta=2.3$, where two pre-collapsed snaking structures appear. The blue curve $\Sigma$ will eventually lead to the collapsed snaking shown in Fig.~\ref{fig1}(c) once the top part of the diagram connects with TI$_2^+$. The violet curve $\Sigma'$ corresponds to other type of mixed states, separated by $L/2$, which emanate from the small amplitude states with a hole in their center. Here, for simplicity we do not show such states and refer the interested reader to other works \cite{parra-rivas_dark_2016,parra-rivas_formation_2020}.
 
This complex re-connection takes place in a cascade of necking bifurcation occurring in other parameter regimes. At low detuning, for example, a similar process leads to the destruction of the standard homoclinic snaking $\Gamma_0^b$ and the consequent formation of isolas $\Theta^b_{1,2,3,4}$ [see Fig.~\ref{fig2}(b)]. In this case, the bifurcation involved in such transition is N$_1^b$, and is marked in the close-up view of Fig.~\ref{fig2}(d). 

Although, we have just focused on the reorganization of the bottom part of the bifurcation diagrams shown in Fig.~\ref{fig2}, similar bifurcation events occurs regarding the top part of such diagrams. This leads to an even more complex scenario, whose complete unfolding requires further study.

\section{Discussions and conclusions}\label{sec:6}
In one-dimensional extended system, the formation of LSs and their organization 
in the parameter space is dictated, in general, by the interaction of front waves between extended stable states which can be of different nature, and may involve two or even more different states.

When two stable states coexist, two main bifurcation organization appears: the standard and collapsed homoclinic snakings. These structures have been widely studied in different fields ranging from photonics to population dynamics \cite{gomila_bifurcation_2007,parra-rivas_origin_2021,zelnik_implications_2018,al_saadi_unified_2021,schmidt_bumps_2020,thompson_advances_2015}. Nowadays, one can say that both structures are quite well understood separately, and a strong mathematical framework supports this fact \cite{woods_heteroclinic_1999,knobloch_homoclinic_2005,gomila_bifurcation_2007,kozyreff_asymptotics_2006,dean_exponential_2011,makrides_existence_2019}. 
%Things are quite different when multiple coexisting extended states coexist. In this case, more complex hybrid fronts appear, leading to important implications regarding the formation of new exotic LSs \cite{zelnik_implications_2018}.

 %This transiton were analyzed in detail in the in gradient systems \cite{parra-rivas_organization_2023}. 

In this article, we have analyzed the formation of LSs in Kerr passive cavities in tristable regimes where two uniform states coexist with an extended pattern. We have referred to this scenario as {\it uniform-pattern-uniform} tristable regime. For this analysis, we have considered the LLE which is traditionally used for understanding the spatiotemporal dynamics of light pulses in Kerr dispersive passive cavities \cite{haelterman_dissipative_1992,parra-rivas_origin_2021}. Here, such tristable regime emerges when the normalized dispersion coefficients are set to $d_2=-1$ and $d_4=1$. This configuration was tackled, although only superficially, in previous works, where the existence of standard homoclinic snaking for dark and bright LSs was predicted \cite{tlidi_high-order_2010}. Our present work, shows the existence of collapsed snaking and the complex transition between the former and the latter. Performing a systematic and detailed bifurcation analysis we show that this transition is mediated by a cascade of codim-2 {\it necking bifurcations} which are related to the emergence of new hybrid states.

This work, which focuses on a non-variational system, shows a similar bifurcation organization than those presented in variational (i.e., gradient-like) context \cite{parra-rivas_organization_2023} demonstrating that such transition is generic in 1D extended systems which are far from thermodynamics equilibrium.

Necking bifurcations have been also demonstrated in the context of fluid dynamics, particularly in two-dimensional Rayleigh–Bénard convection problems \cite{prat_12_2002}.
Besides, such bifurcations are responsible of the bifurcation scenario found in other pattern forming systems including plant population dynamics \cite{zelnik_implications_2018, al_saadi_transitions_2023}.

Regarding the experimental demonstration of such states in a cavity optics context, the dispersion regime discussed here could be reached using advanced dispersion engineering techniques as those based, for example, on an intracavity pulse shaper  \cite{runge_pure-quartic_2020}. This technique, initially proposed in the context of soliton lasers, opens astoning posibilities and new avenues for LSs formation and control.

Regarding higher extended dimensions, it is unknown how this scenario could be generalized. This problem could be tackled in the case of radially symmetric states by continuing homotopically, in the system dimension, the LSs in 1D to 2D and 3D. Some studies of this kind have been performed regarding standard homoclinic related states in the Swift-Hohenberg equation \cite{mccalla_snaking_2010} and spike-like solitons in passive cavities with parabolic potentials \cite{sun_2_2023}. However, the transition shown here is completely unknown in those cases. We will address these question in future investigations. 
%Some of the future work on this topic should definitely tackle this issue. 

\bibliographystyle{ieeetr}
\bibliography{Refs}

\begin{thebibliography}{10}

\bibitem{pomeau_front_1986}
Y.~Pomeau, ``Front motion, metastability and subcritical bifurcations in
  hydrodynamics,'' {\em Physica D: Nonlinear Phenomena}, vol.~23, pp.~3--11,
  Dec. 1986.

\bibitem{coullet_nature_1987}
P.~Coullet, C.~Elphick, and D.~Repaux, ``Nature of spatial chaos,'' {\em
  Physical Review Letters}, vol.~58, pp.~431--434, Feb. 1987.

\bibitem{thual_localized_1988}
O.~Thual and S.~Fauve, ``Localized structures generated by subcritical
  instabilities,'' {\em Journal de Physique}, vol.~49, no.~11, pp.~1829--1833,
  1988.

\bibitem{coullet_localized_2002}
P.~Coullet, ``Localized patterns and fronts in nonequilibrium systems,'' {\em
  International Journal of Bifurcation and Chaos}, vol.~12, pp.~2445--2457,
  Nov. 2002.

\bibitem{woods_heteroclinic_1999}
P.~D. Woods and A.~R. Champneys, ``Heteroclinic tangles and homoclinic snaking
  in the unfolding of a degenerate reversible {Hamiltonian}–{Hopf}
  bifurcation,'' {\em Physica D: Nonlinear Phenomena}, vol.~129, pp.~147--170,
  May 1999.

\bibitem{knobloch_homoclinic_2005}
J.~Knobloch and T.~Wagenknecht, ``Homoclinic snaking near a heteroclinic cycle
  in reversible systems,'' {\em Physica D: Nonlinear Phenomena}, vol.~206,
  pp.~82--93, June 2005.

\bibitem{parra-rivas_dark_2016}
P.~Parra-Rivas, E.~Knobloch, D.~Gomila, and L.~Gelens, ``Dark solitons in the
  {Lugiato}-{Lefever} equation with normal dispersion,'' {\em Physical Review
  A}, vol.~93, p.~063839, June 2016.

\bibitem{gomila_bifurcation_2007}
D.~Gomila, A.~J. Scroggie, and W.~J. Firth, ``Bifurcation structure of
  dissipative solitons,'' {\em Physica D: Nonlinear Phenomena}, vol.~227,
  pp.~70--77, Mar. 2007.

\bibitem{parra-rivas_bifurcation_2018}
P.~Parra-Rivas, D.~Gomila, L.~Gelens, and E.~Knobloch, ``Bifurcation structure
  of localized states in the {Lugiato}-{Lefever} equation with anomalous
  dispersion,'' {\em Physical Review E}, vol.~97, p.~042204, Apr. 2018.

\bibitem{parra-rivas_localized_2019}
P.~Parra-Rivas, L.~Gelens, and F.~Leo, ``Localized structures in dispersive and
  doubly resonant optical parametric oscillators,'' {\em Physical Review E},
  vol.~100, p.~032219, Sept. 2019.

\bibitem{parra-rivas_parametric_2020}
P.~Parra-Rivas, C.~Mas-Arabí, and F.~Leo, ``Parametric localized patterns and
  breathers in dispersive quadratic cavities,'' {\em Physical Review A},
  vol.~101, p.~063817, June 2020.

\bibitem{parra-rivas_origin_2021}
P.~Parra-Rivas, E.~Knobloch, L.~Gelens, and D.~Gomila, ``Origin, bifurcation
  structure and stability of localized states in {Kerr} dispersive optical
  cavities,'' {\em IMA Journal of Applied Mathematics}, vol.~86, pp.~856--895,
  Oct. 2021.

\bibitem{al_saadi_unified_2021}
F.~Al~Saadi and A.~Champneys, ``Unified framework for localized patterns in
  reaction–diffusion systems; the {Gray}–{Scott} and {Gierer}–{Meinhardt}
  cases,'' {\em Philosophical Transactions of the Royal Society A:
  Mathematical, Physical and Engineering Sciences}, vol.~379, p.~20200277, Dec.
  2021.
\newblock Publisher: Royal Society.

\bibitem{zelnik_implications_2018}
Y.~R. Zelnik, P.~Gandhi, E.~Knobloch, and E.~Meron, ``Implications of
  tristability in pattern-forming ecosystems,'' {\em Chaos: An
  Interdisciplinary Journal of Nonlinear Science}, vol.~28, p.~033609, Mar.
  2018.

\bibitem{al_saadi_transitions_2023}
F.~Al~Saadi and P.~Parra-Rivas, ``Transitions between dissipative localized
  structures in the simplified {Gilad}–{Meron} model for dryland plant
  ecology,'' {\em Chaos: An Interdisciplinary Journal of Nonlinear Science},
  vol.~33, p.~033129, Mar. 2023.

\bibitem{schmidt_bumps_2020}
H.~Schmidt and D.~Avitabile, ``Bumps and oscillons in networks of spiking
  neurons,'' {\em Chaos: An Interdisciplinary Journal of Nonlinear Science},
  vol.~30, p.~033133, Mar. 2020.

\bibitem{parra-rivas_organization_2023}
P.~Parra-Rivas, A.~R. Champneys, F.~A. Saadi, D.~Gomila, and E.~Knobloch,
  ``Organization of {Spatially} {Localized} {Structures} near a
  {Codimension}-{Three} {Cusp}-{Turing} {Bifurcation},'' {\em SIAM Journal on
  Applied Dynamical Systems}, pp.~2693--2731, Dec. 2023.
\newblock Publisher: Society for Industrial and Applied Mathematics.

\bibitem{cross_pattern_1993}
M.~C. Cross and P.~C. Hohenberg, ``Pattern formation outside of equilibrium,''
  {\em Reviews of Modern Physics}, vol.~65, pp.~851--1112, July 1993.

\bibitem{lugiato_spatial_1987}
L.~A. Lugiato and R.~Lefever, ``Spatial {Dissipative} {Structures} in {Passive}
  {Optical} {Systems},'' {\em Physical Review Letters}, vol.~58,
  pp.~2209--2211, May 1987.

\bibitem{haelterman_dissipative_1992}
M.~Haelterman, S.~Trillo, and S.~Wabnitz, ``Dissipative modulation instability
  in a nonlinear dispersive ring cavity,'' {\em Optics Communications},
  vol.~91, pp.~401--407, Aug. 1992.

\bibitem{rowley_self-emergence_2022}
M.~Rowley, P.-H. Hanzard, A.~Cutrona, H.~Bao, S.~T. Chu, B.~E. Little,
  R.~Morandotti, D.~J. Moss, G.-L. Oppo, J.~S. Totero~Gongora, M.~Peccianti,
  and A.~Pasquazi, ``Self-emergence of robust solitons in a microcavity,'' {\em
  Nature}, vol.~608, pp.~303--309, Aug. 2022.
\newblock Number: 7922 Publisher: Nature Publishing Group.

\bibitem{lottes_excitation_2021}
J.~Lottes, G.~Biondini, and S.~Trillo, ``Excitation of switching waves in
  normally dispersive {Kerr} cavities,'' {\em Optics Letters}, vol.~46,
  pp.~2481--2484, May 2021.
\newblock Publisher: Optica Publishing Group.

\bibitem{chembo_spatiotemporal_2013}
Y.~K. Chembo and C.~R. Menyuk, ``Spatiotemporal {Lugiato}-{Lefever} formalism
  for {Kerr}-comb generation in whispering-gallery-mode resonators,'' {\em
  Physical Review A}, vol.~87, p.~053852, May 2013.

\bibitem{tlidi_high-order_2010}
M.~Tlidi and L.~Gelens, ``High-order dispersion stabilizes dark dissipative
  solitons in all-fiber cavities,'' {\em Optics Letters}, vol.~35,
  pp.~306--308, Feb. 2010.

\bibitem{akakpo_emergence_2023}
E.~K. Akakpo, M.~Haelterman, F.~Leo, and P.~Parra-Rivas, ``Emergence of
  collapsed snaking related dark and bright {Kerr} dissipative solitons with
  quartic-quadratic dispersion,'' {\em Physical Review E}, vol.~108, p.~014203,
  July 2023.
\newblock Publisher: American Physical Society.

\bibitem{parra-rivas_quartic_2022}
P.~Parra-Rivas, S.~Hetzel, Y.~V. Kartashov, P.~F.~d. Córdoba, J.~A. Conejero,
  A.~Aceves, and C.~Milián, ``Quartic {Kerr} cavity combs: bright and dark
  solitons,'' {\em Optics Letters}, vol.~47, pp.~2438--2441, May 2022.
\newblock Publisher: Optica Publishing Group.

\bibitem{doedel_numerical_1991}
E.~Doedel, H.~B. Keller, and J.~P. Kernevez, ``Numerical analysis and control
  of bifurcation problems (ii): bifurcation in infinite dimensions,'' {\em
  International Journal of Bifurcation and Chaos}, vol.~01, pp.~745--772, Dec.
  1991.

\bibitem{doedel_numerical_1991-1}
E.~Doedel, H.~B. Keller, and J.~P. Kernevez, ``Numerical analysis and control
  of bifurcation problems (i): bifurcation in finite dimensions,'' {\em
  International Journal of Bifurcation and Chaos}, vol.~01, pp.~493--520, Sept.
  1991.

\bibitem{Doedel2009}
E.~J. Doedel, A.~R. Champneys, T.~F. Fairgrieve, Y.~A. Kuznetsov, B.~Sandstede,
  and X.~Wang, ``{AUTO-07p: Software for continuation and bifurcation problems
  in ordinary differential equations},'' {\em Department of Computer Science,
  Concordia University, Montreal}, 2007.

\bibitem{wiggins_introduction_2003}
S.~Wiggins, {\em Introduction to {Applied} {Nonlinear} {Dynamical} {Systems}
  and {Chaos}}.
\newblock Texts in {Applied} {Mathematics}, New York: Springer-Verlag, 2~ed.,
  2003.

\bibitem{scroggie_pattern_1994}
A.~J. Scroggie, W.~J. Firth, G.~S. McDonald, M.~Tlidi, R.~Lefever, and L.~A.
  Lugiato, ``Pattern formation in a passive {Kerr} cavity,'' {\em Chaos,
  Solitons \& Fractals}, vol.~4, pp.~1323--1354, Aug. 1994.

\bibitem{makrides_existence_2019}
E.~Makrides and B.~Sandstede, ``Existence and stability of spatially localized
  patterns,'' {\em Journal of Differential Equations}, vol.~266,
  pp.~1073--1120, Jan. 2019.

\bibitem{burke_snakes_2007}
J.~Burke and E.~Knobloch, ``Snakes and ladders: {Localized} states in the
  {Swift}–{Hohenberg} equation,'' {\em Physics Letters A}, vol.~360,
  pp.~681--688, Jan. 2007.

\bibitem{parra-rivas_origin_2016}
P.~Parra-Rivas, D.~Gomila, E.~Knobloch, S.~Coen, and L.~Gelens, ``Origin and
  stability of dark pulse {Kerr} combs in normal dispersion resonators,'' {\em
  Optics Letters}, vol.~41, pp.~2402--2405, June 2016.

\bibitem{prat_12_2002}
J.~Prat, I.~Mercader, and E.~Knobloch, ``The 1:2 mode interaction in
  rayleigh–bénard convection with and without boussinesq symmetry,'' {\em
  International Journal of Bifurcation and Chaos}, vol.~12, pp.~281--308, Feb.
  2002.

\bibitem{parra-rivas_formation_2020}
P.~Parra-Rivas and C.~Fernandez-Oto, ``Formation of localized states in dryland
  vegetation: {Bifurcation} structure and stability,'' {\em Physical Review E},
  vol.~101, p.~052214, May 2020.

\bibitem{thompson_advances_2015}
J.~M.~T. Thompson, ``Advances in {Shell} {Buckling}: {Theory} and
  {Experiments},'' {\em International Journal of Bifurcation and Chaos}, Feb.
  2015.
\newblock Publisher: World Scientific Publishing Company.

\bibitem{kozyreff_asymptotics_2006}
G.~Kozyreff and S.~J. Chapman, ``Asymptotics of {Large} {Bound} {States} of
  {Localized} {Structures},'' {\em Physical Review Letters}, vol.~97,
  p.~044502, July 2006.

\bibitem{dean_exponential_2011}
A.~D. Dean, P.~C. Matthews, S.~M. Cox, and J.~R. King, ``Exponential
  asymptotics of homoclinic snaking,'' {\em Nonlinearity}, vol.~24, p.~3323,
  Oct. 2011.

\bibitem{runge_pure-quartic_2020}
A.~F.~J. Runge, D.~D. Hudson, K.~K.~K. Tam, C.~M. de~Sterke, and
  A.~Blanco-Redondo, ``The pure-quartic soliton laser,'' {\em Nature
  Photonics}, vol.~14, pp.~492--497, Aug. 2020.
\newblock Number: 8 Publisher: Nature Publishing Group.

\bibitem{mccalla_snaking_2010}
S.~McCalla and B.~Sandstede, ``Snaking of radial solutions of the
  multi-dimensional {Swift}–{Hohenberg} equation: {A} numerical study,'' {\em
  Physica D: Nonlinear Phenomena}, vol.~239, pp.~1581--1592, Aug. 2010.

\bibitem{sun_2_2023}
Y.~Sun, P.~Parra-Rivas, M.~Zitelli, F.~Mangini, M.~Ferraro, and S.~Wabnitz, ``2
  - {An} introduction to guided-wave nonlinear ultrafast photonics,'' in {\em
  Advances in {Nonlinear} {Photonics}} (G.~C. Righini and L.~Sirleto, eds.),
  Woodhead {Publishing} {Series} in {Electronic} and {Optical} {Materials},
  pp.~27--55, Woodhead Publishing, Jan. 2023.

\end{thebibliography}
\end{document}